\documentclass[twocolumn,english,aps,prl,10pt,showpacs]{revtex4-1}
\usepackage[UKenglish]{babel}
\usepackage[colorlinks=true,urlcolor=blue,citecolor=blue,linkcolor=blue]{hyperref}
\usepackage[sort&compress]{natbib}

\usepackage{varioref,amssymb,textcomp,bm,amsfonts}
\usepackage{amsmath}
\usepackage[pdftex]{graphicx}
\usepackage{keyval, everysel, ragged2e}
\usepackage{tabularx}

\usepackage{subfig}

\newcommand{\ua}{\uparrow}
\newcommand{\da}{\downarrow}

\graphicspath{{figures/}}

\begin{document}


\captionsetup{font={footnotesize},justification=raggedright}

\title{Ferromagnetism of the Repulsive Atomic Fermi Gas: three-body
  recombination and domain formation}

\author{Ilia Zintchenko$^{1}$, Lei Wang$^{1,2}$ and Matthias Troyer$^1$}

\affiliation{$^{1}$Theoretische Physik, ETH Zurich, 8093 Zurich,
  Switzerland}

\affiliation{$^{2}$Beijing National Lab for Condensed Matter Physics
  and Institute of Physics, Chinese Academy of Sciences, Beijing
  100190, China}

\begin{abstract}
  The simplest model for itinerant ferromagnetism, the Stoner model,
  has so far eluded experimental observation in repulsive ultracold
  fermions due to rapid three-body recombination at large scattering
  lengths. Here we show that a ferromagnetic phase can be stabilised
  by imposing a moderate optical lattice. The reduced kinetic energy
  drop upon formation of a polarized phase in an optical lattice
  extends the ferromagnetic phase to smaller scattering lengths where
  three-body recombination is small enough to permit experimental
  detection of the phase. We also show, using time dependent density
  functional theory, that in such a setup ferromagnetic domains emerge
  rapidly from a paramagnetic initial state.
\end{abstract}
\maketitle

The ground-state of a repulsive gas of fermions with contact
interaction was first predicted by
Stoner~\cite{doi:10.1080/14786443309462241} in 1933 to be
ferromagnetic and a precise value for the critical interaction
strength in a homogeneous system was recently obtained with diffusion
Monte Carlo simulations ~\cite{QMCpaper, Pilati:2010p20001,
  Chang:2011p34450}. While ultracold fermionic gases should provide a
controlled environment to study this phenomenon, the instability of a
strongly repulsive gas towards molecule formation has so far prevented
experimental realization of this phase. Repulsive fermions on the
positive side of the Feshbach resonance live on the meta-stable
repulsive branch~\cite{RevModPhys.80.1215}. When three atoms, one with
the opposite spin of the other two, come close to each other two atoms
with opposite spin will form bosonic molecules and the other one
carries the binding energy away. The rate of such process increases
rapidly with scattering length. The lifetime of the gas is therefore
largely governed by the interaction strength and the spatial overlap
between the two spin species.

A recent experiment~\cite{Jo:2009p10288} presented evidence for a
possible ferromagnetic state formed after rapid increase of the
scattering length. The nature of this phase has, however, been
questioned~\cite{PhysRevA.80.051605, PhysRevLett.108.240404,
  PhysRevLett.106.050402} as the peaks in kinetic energy, cloud size
and loss rate observed in~\cite{Jo:2009p10288} are only indirect
evidence for ferromagnetic domains~\cite{PhysRevA.80.051605}, and it
has been shown that molecule formation is dominant at large
interaction strengths~\cite{PhysRevLett.106.050402,
  PhysRevLett.108.240404}. Several papers have proposed to reduce the
recombination rate by using a polar molecular gas with dipole
interactions and positive scattering range~\cite{arXiv5, arXiv6},
narrower Feshbach resonances~\cite{Hazlett:2012p56056,
  PhysRevLett.106.050402, PhysRevLett.108.240404, Lee:2012p58217} and
fermions with unequal mass~\cite{vonKeyserlingk:2011p33313,
  Cui:2012p60813}. Although these approaches might prove promising in
future experiments, they all change the microscopic physics of the
Stoner model.

Here we suggest new strategies which achieve the same goal of
stabilising the ferromagnetic phase, yet preserve the microscopic
physics and thus pave the way towards experimental realization of
Stoner ferromagnetism. Firstly, the lifetime of the system can be
increased by reducing the overlap volume between polarized domains
with different spins. This can be achieved using elongated traps with
larger aspect ratios, but turns out to be insufficient to stabilise
the phase by itself. A more effective way is to reduce the critical
scattering length, at which ferromagnetism sets in, by imposing a
shallow optical lattice. The ferromagnetic transition is determined by
a competition between the loss of kinetic energy and gain of
interaction energy. Since the optical lattice reduces the kinetic
energy scale, the ferromagnetic state is stabilised. A similar effect
is found in flat band
ferromagnetism~\cite{PhysRevLett.62.1201,Mielke91,PhysRevLett.69.1608}

\begin{figure}
\centering
  \includegraphics[width=0.9\columnwidth]{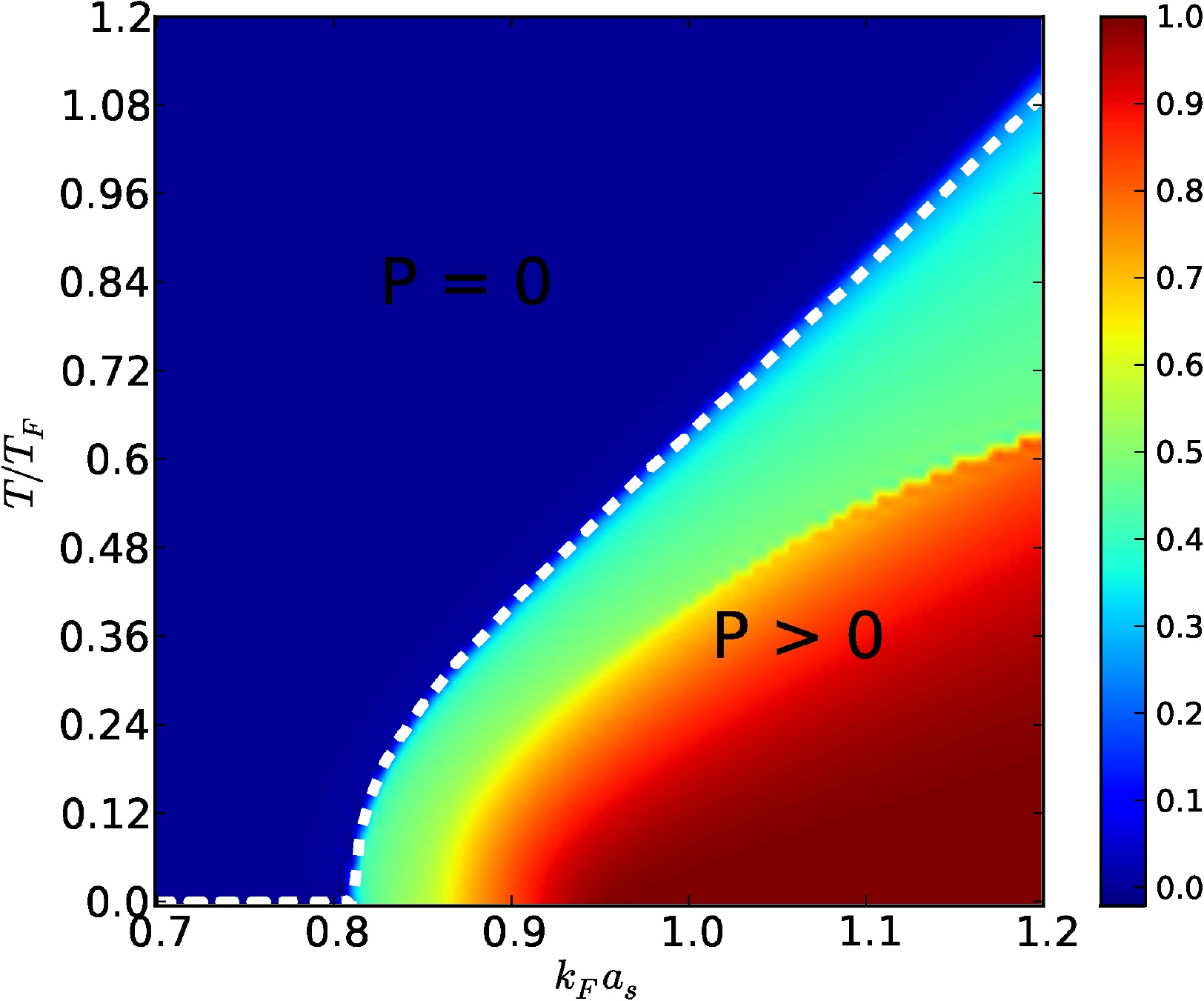}
  \caption{Phase diagram of the homogeneous repulsive Fermi gas as a
    function of temperature $T/T_F$ and interaction strength $k_F
    a_s$. The white dashed line indicates the paramagnet-ferromagnetic
    phase transition and the colour scale the polarization $P = (n_\ua
    - n_\da) / (n_\ua + n_\da)$.}
\label{fig:phasediag_temp}
\end{figure}

To obtain quantitative results for the phase diagrams we use Kohn-Sham
density functional theory~\cite{PhysRev.136.B864,PhysRev.140.A1133}
where the exchange-correlation energy is treated within a local spin
density approximation (LSDA) which has been widely used for materials
simulations and more recently for ultracold atomic
gases~\cite{Bulgac10062011, PhysRevC.84.051309, PhysRevA.83.032503,
  Ma:2012p60558}. The LSDA exchange-correlation functional is obtained
by solving a uniform system at zero temperature with diffusion
Monte-Carlo~\cite{Pilati:2010p20001, Chang:2011p34450}, where
interactions between fermionic atoms with opposite spin are modelled
by a hard-sphere potential with scattering length $a_s$.

Before discussing our proposal to stabilise the ferromagnetic phase we
investigate whether thermal fluctuations, which can significantly
affect of the stability of the ferromagnetic
phase~\cite{PhysRevLett.106.080402}, might be responsible for the
absence of a stable ferromagnetic phase in experiments. To quantify
the effect of non-zero temperature we employ finite-temperature
density functional theory with a zero temperature exchange-correlation
correction~\cite{Stoitsov1988121, 0953-8984-14-40-302,
  PhysRev.137.A1441, Gupta1982259}. The resulting phase diagram,
presented in Fig. \ref{fig:phasediag_temp}, is in general agreement
with previous results~\cite{PhysRevLett.95.230403,
  PhysRevA.83.053635}.  We observe almost full polarization in the
currently experimentally accessible temperature regime $T \sim 0.25
T_F$~\cite{Lee:2012p58217, PhysRevLett.108.240404} and thermal
fluctuations are thus not the dominant mechanism destabilising the
ferromagnetic phase.

\begin{figure}
  \centering
  \includegraphics[width=1.0\columnwidth]{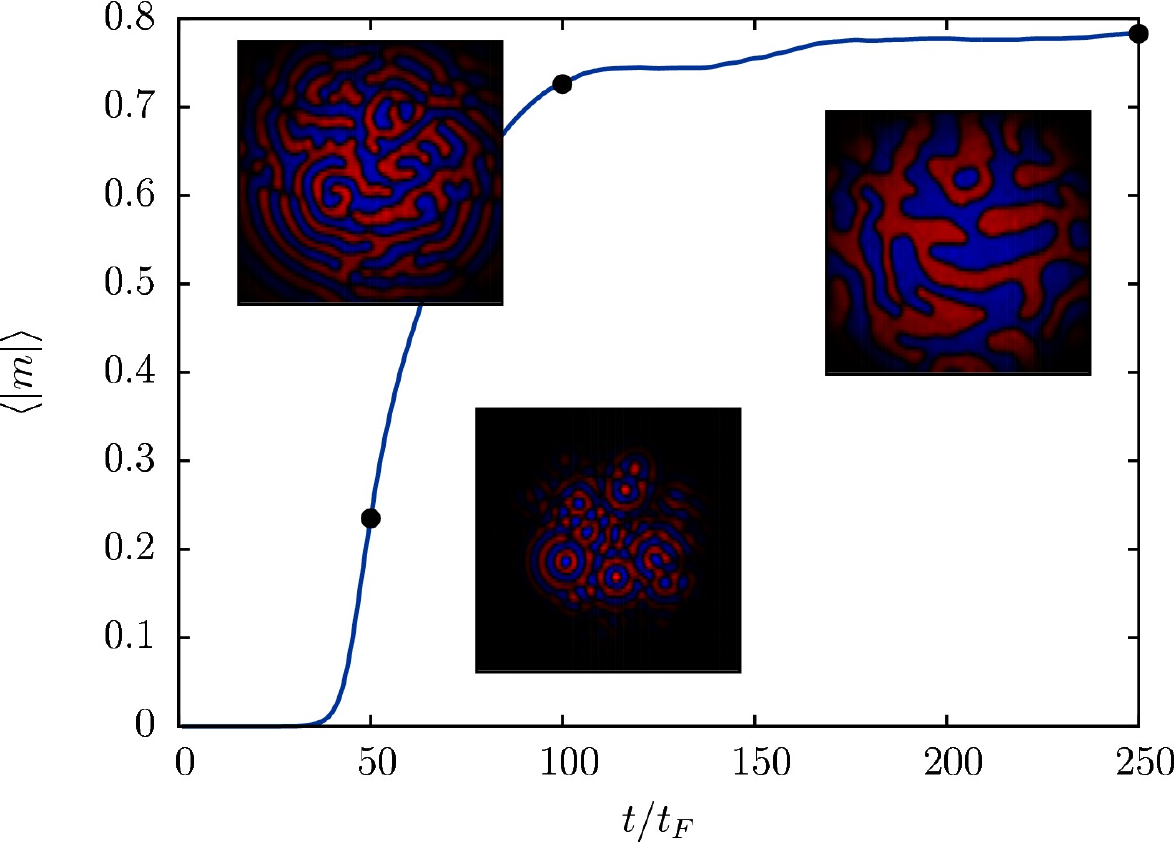}
  \caption{Total magnetisation $\langle |m| \rangle$ in pancake shaped
    harmonic confinement for $N = 440$. Starting from a paramagnetic
    phase the system evolves in the presence of thermal noise for $t
    \in [0;250] t_F$. Topology of ferromagnetic domains at $t = \{ 50,
    150, 250\} t_F$ is shown in the insets. As the gas remains
    unpolarised away from the center, the maximum magnetisation is
    $\sim 0.8$.}
  \label{fig:dynamics}
\end{figure}

An equally important question is that of the time scale over which
ferromagnetic domains form from an initially paramagnetic state, which
should be within the capabilities of current experimental setups for
the observation of ferromagnetic domains to remain plausible. We
address this issue within the time-dependent Kohn-Sham
formalism~\cite{PhysRevLett.52.997} ignoring the effect of three-body
recombination. The system is evolved in pancake shaped harmonic
confinement in the presence of thermal noise after a quench of the
interaction strength $k_F^0 a_s$ to $1.2$. Already after $t \sim 250
t_F$ ferromagnetic domains with a feature size $\sim 10 k_F^{-1}$ form
(Fig. \ref{fig:dynamics}), which is within the resolution currently
achievable in the lab. The ferromagnetic phase is also significantly
more stable with a total recombination rate reduced by a factor $\sim
3$ relative to the initial paramagnetic gas. In the experiment
~\cite{PhysRevLett.108.240404}, more than half of the particles remain
in the meta-stable repulsive branch after $250 t_F$. However,
ferromagnetic domains were not observed indicating that three-body
recombination has more significant effects beyond reducing the
fraction of meta-stable fermions in the system.

We thus focus on three-body recombination and determine its effect on
the stability of the ferromagnetic phase. The loss rate $\Gamma =
-\partial_t N / N$ ($N$ is the total number of particles in the
system) is inversely proportional to the lifetime of the system and
can be computed as~\cite{Petrov:2003p54744}
\begin{equation}
  \Gamma \sim {a_s^3}\sum_{\sigma}\int d\bm{r}
  \int_{|\bm{r}^\prime-\bm{r}|<a_s} d\bm{r}^\prime
  \varepsilon_{F}(\bm{r}) g_{\bar{\sigma}\sigma\sigma} (\bm{r},\bm{r},
  \bm{r}^{\prime})
  \label{eqn:loss_rate}
\end{equation}
where $\varepsilon_{F}(\bm{r})=\hbar^{2}(3\pi^2 (n_\ua +
n_\da))^{2/3}/2m$ is the local Fermi energy, $n_\sigma (\bm{r})$ is
the local density of each spin, $m$ the atom mass and
$g_{\bar{\sigma}\sigma\sigma} (\bm{r},\bm{r},\bm{r}^{\prime}) =
\langle \hat{\psi}^\dagger_{{\bar{\sigma}}} (\bm{r})
\hat{\psi}^\dagger_{{\sigma}} (\bm{r}) \hat{\psi}^\dagger_{{\sigma}}
(\bm{r}^\prime) \hat{\psi}_{{\sigma}} (\bm{r}^\prime)
\hat{\psi}_{{\sigma}} (\bm{r}) \hat{\psi}_{{\bar{\sigma}}} (\bm{r})
\rangle$ is the three-body correlation function which gives the
probability of finding three particles with spin
$\bar{\sigma}\sigma\sigma$~\cite{PhysRevLett.77.2921} at locations
$\bm{r} \bm{r} \bm{r}^\prime$ respectively. Here $\bar{\sigma}$ refers
to the opposite spin of $\sigma$. In this way the total loss rate is
in units of energy and the microscopic contribution, represented by
the pre-factor $a_s^3$ together with the integration over
$\bm{r}^{\prime}$, is decoupled from the many-body background given by
the correlation function.  The three-particle correlation function is
commonly~\cite{PhysRevLett.95.230403, Conduit:2011p55199,
  PhysRevA.80.013607} treated phenomenologically as
$g_{\bar{\sigma}\sigma\sigma}^0 (\bm{r}, \bm{r},\bm{r}^{\prime}) =
n_{\bar{\sigma}}(\bm{r}) n_{\sigma}(\bm{r})
n_{\sigma}(\bm{r}^{\prime})$, where the density is assumed to be
constant within the integral over $\bm{r}^\prime$. This form
disregards quantum mechanical corrections due to exchange effects and
violates the Pauli principle. Within the Kohn-Sham formalism a more
accurate representation is
\begin{equation}
  g_{\bar{\sigma}\sigma\sigma} (\bm{r}, \bm{r},\bm{r}^{\prime}) =
  g_{\bar{\sigma}\sigma\sigma}^0 (\bm{r}, \bm{r},\bm{r}^{\prime}) -
  n_{\bar{\sigma}}(\bm{r}) |\rho_{\sigma}(\bm{r},\bm{r}^{\prime})|^{2}
\label{correlator}
\end{equation}
where $\rho_{\sigma}(\bm{r},\bm{r}^{\prime})$ is the one-body density
matrix. The Pauli principle is restored in this formulation as
$g_{\bar{\sigma}\sigma\sigma} (\bm{r}, \bm{r},\bm{r})\equiv 0$, thus
leading to a better estimate of three-body losses.

\begin{figure}
  \centering
  \subfloat[Total density]{\includegraphics[width=1.0\columnwidth]{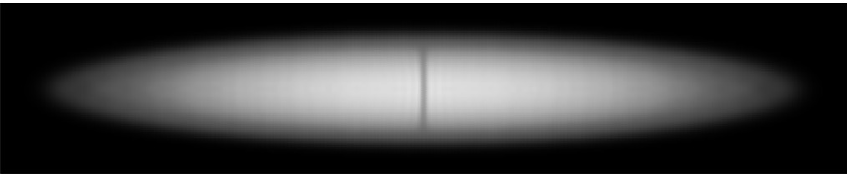}\label{fig:density_pos}}\\
  \subfloat[Magnetisation]{\includegraphics[width=1.0\columnwidth]{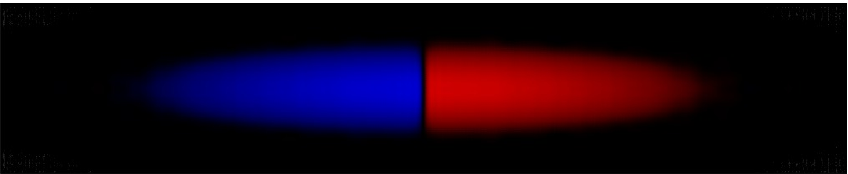}\label{fig:mag_pos}}\\
  \subfloat[Recombination rate]{\includegraphics[width=1.0\columnwidth]{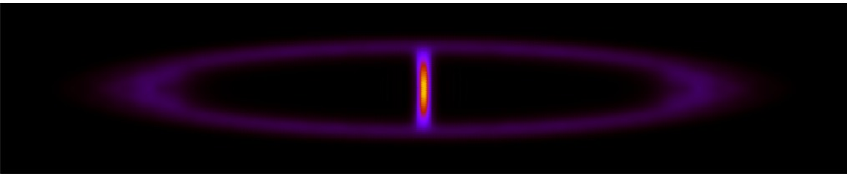}\label{fig:gamma_pos}}
  \caption{Total density $n_\ua + n_\da$, magnetisation $n_\ua -
    n_\da$ and loss rate $\Gamma$ in harmonic confinement for the
    aspect ratio $\lambda = 5$, $k_F^0 a_s \sim 1.4$ and $N_\ua =
    N_\da = 969$. The $z$ and $r$ axes are along the horizontal and
    vertical directions respectively.}
  \label{fig:pos_trap}
\end{figure}

As molecule formation in the gas is a three-body process and requires
both species of fermions to be close to each other, the recombination
rate (\ref{eqn:loss_rate}) is determined by the total volume where the
two spin species are mixed. In a polarised system this amounts to the
overlap between ferromagnetic domains at their boundaries, whose
structure strongly depends on the external confining potential. To
further investigate this effect, we impose a cigar shaped harmonic
potential, which has been used in a number of recent
experiments~\cite{PhysRevLett.108.240404, Lee:2012p58217}. With
cylindrical symmetry around the $z$-axis the potential is in the form
$V_{HO}(r,z) = 1/2(\omega_r^2 r^2 + \omega_z z^2)$ with aspect ratio
$\lambda = \omega_r / \omega_z$. We will consider the spin-balanced
case $N_\ua = N_\da = N/2$ and normalise the loss rate $\Gamma$ by the
critical recombination rate $\Gamma_c$, above which the system is
unstable. This value can be estimated from the experimental
observation of a maximum tolerable $k_F a_s\sim 0.4$ for $^{6}$Li
atoms in an optical dipole trap~\cite{Lee:2012p58217}.

With equal number of spin-up and down particles, the ferromagnetic
ground state exhibits a domain wall across the middle of the trap, as
shown in Fig. \ref{fig:mag_pos}. Due to repulsion between the two
polarized domains there is a drop in total density along this region
with a width $\sim 0.5 k_F^{-1}$ (see Fig. \ref{fig:density_pos}),
which is, however, below the resolution of current
experiments~\cite{PhysRevLett.108.240404}. As shown in
Fig. \ref{fig:gamma_pos}), particle loss occurs predominantly at the
domain wall in the center of the trap and in the low-density halo on
the surface of the clouds where the gas remains unpolarised due to low
densities.

\begin{figure}
  \centering
  \includegraphics[width=0.95\columnwidth]{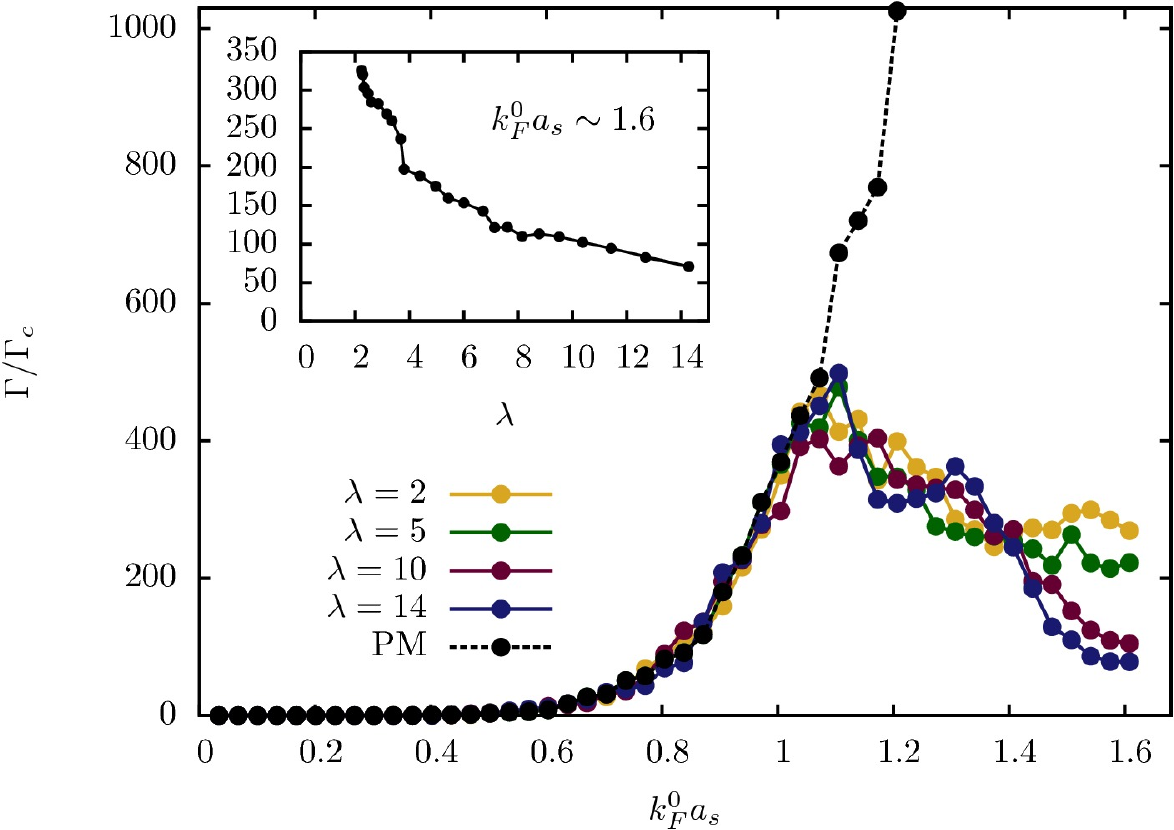}
  \caption{Normalised recombination rate $\Gamma / \Gamma_c$ in
    harmonic confinement for aspect ratios $\lambda = \{2, 5, 10,
    14\}$ and $N_\ua = N_\da = 560$ at different interaction
    strengths. \emph{Inset:} dependence of $\Gamma / \Gamma_c$ on the
    aspect ratio for $k_F^0 a_s \sim 1.6$. From simulations of
    different system sizes we checked that the results are not
    sensitive to the number of atoms in the system.}
  %% The system is partially polarised for $k_F^0 a_s \in
  %% [1.1; 1.4]$ and fully polarised for $k_F^0 a_s >
  %% 1.4$. 
  \label{fig:gamma_trap_lambda}
\end{figure}

One can thus expect that larger aspect ratios may reduce losses due to
a smaller area of the domain wall. To investigate this quantitatively
we present the results of our calculations for the recombination rate
in Fig. \ref{fig:gamma_trap_lambda} as a function of interaction
strength $k_Fa_a$. Starting from a non-interacting gas in the
paramagnetic ground state, the recombination rate increases until
phase separation due to ferromagnetism sets in and a maximum is
reached at $k_F a_s \sim 1.1$, after which total losses are again
reduced. A similar dependence of the loss rate on scattering length
was observed in the experiment~\cite{Jo:2009p10288} and theoretical
studies~\cite{PhysRevA.80.013607, 1367-2630-13-5-055003}. As the trap
is compressed around the $z$-axis at constant volume $\bar{\omega} =
(\omega_r^2 \omega_z)^{1/3}$ and system size $N$ (see inset of
Fig. \ref{fig:gamma_trap_lambda}), the aspect ratio increases and the
recombination rate is indeed significantly reduced. However, with a
six times lower recombination rate at $\lambda = 14$ compared to
$\lambda = 2$, the loss rate is still above the critical value
$\Gamma_c$.

\begin{figure}
  \centering
  \includegraphics[width=0.9\columnwidth]{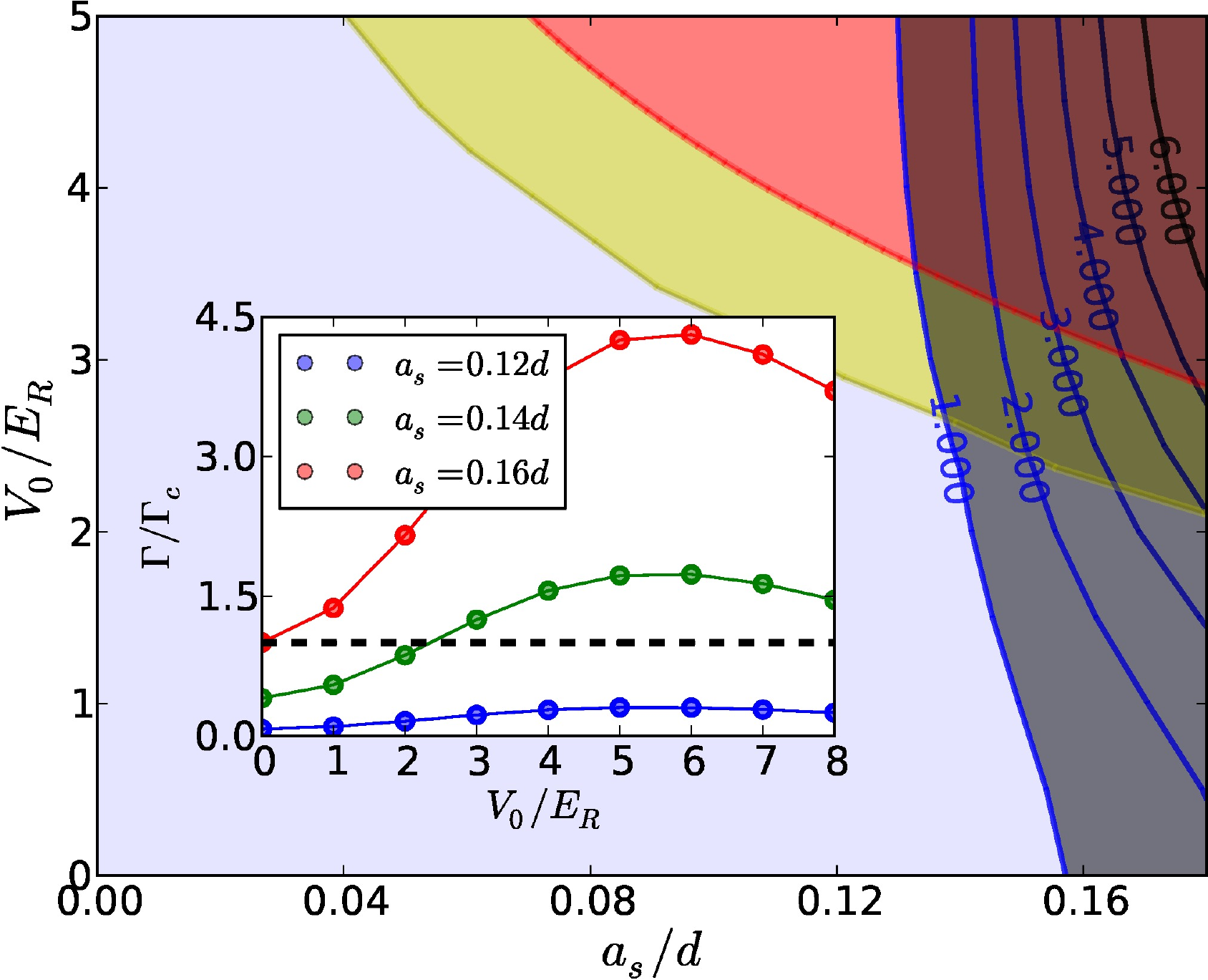}
  \caption{Contours of constant normalised loss rate $\Gamma /
    \Gamma_c$ at quarter filling $\bar{n} = 0.5$ in an optical
    lattice. In the shaded region $\Gamma > \Gamma_c$. Red (yellow)
    region denotes the fully (partially) polarized ferromagnetic
    phase. \emph{Inset:} Normalised loss rates $\Gamma / \Gamma_c$ for
    $a_s = \{0.12, 0.14, 0.16\}$ at different lattice depths. The
    dotted line is $\Gamma_c$. From simulations at different fillings
    bellow $\bar{n} = 1.0$, we checked that the physics remains
    qualitatively the same.}
\label{fig:Gammacontour}
\end{figure}

We now turn to the most promising way of stabilising the
ferromagnetic phase, by reducing the critical scattering length of the
ferromagnetic phase transition. This can be achieved by imposing a
shallow optical lattice~\cite{Ma:2012p60558}, which reduces the
kinetic energy loss at the phase transition and thus favors the
ferromagnetic state. Figure \ref{fig:Gammacontour} shows the phase
diagram at quarter filling ($\bar{n} = 0.5$) in a shallow 3D cubic
optical lattice $V_\mathrm{OL}(\bm{r})= V_{0}\sum_{i=1}^{3}
\sin^{2}(\pi \bm{r}_{i}/d)$, where $V_0$ is in units of recoil energy
$E_{R} = \frac{\hbar^{2}\pi^{2}}{2md^{2}}$ and $d$ is the lattice
constant. At quarter filling $\bar{n} = 0.5$, the maximum tolerable
$k_F a_s \sim 0.4$ corresponds to $a_s = 0.16d$, which is nearly half
the critical scattering length predicted by diffusion Monte
Carlo~\cite{Pilati:2010p20001, Chang:2011p34450}.

We see that a ferromagnetic phase exists at this scattering length for
a lattice depth of about $V_0\sim 2.5 E_R$. However, the loss rate is
likely significantly affected by the presence of a periodic potential.
To address this issue we calculate the recombination rate of the
paramagnetic state in an optical lattice by integrating over the unit
cell Eq.(\ref{eqn:loss_rate}).

The loss rate initially grows as the lattice is ramped up since the
density increases at the center of the unit cell (inset of
Fig. \ref{fig:Gammacontour}). In very deep lattices the loss rate is
reduced when one approaches the regime where the physics is well
described by the single band Hubbard model and triple occupation of a
lattice site is reduced. In Fig. \ref{fig:Gammacontour} we show
contour lines of constant loss rate $\Gamma$ and indicate by grey
shading the regime where we expect three body losses to be above the
critical experimental value. We find that despite the increased
three-body losses in a shallow lattice, the ferromagnetic phase is
stable in a wide region of the phase diagram. Our calculations have
not taken into account the quantum Zeno
effect~\cite{PhysRevLett.87.040402} which will further suppress three
body recombination.

In summary, we have shown that although larger aspect ratios in
harmonic confinement can significantly reduce the total recombination
rate, the gas remains unstable. However, in an optical lattice the
ferromagnetic phase extends down to small enough scattering lengths
where three-body recombination is below the critical
value. Experimental verification of our results will be solid
confirmation of the Stoner model of itinerant ferromagnetism.
%% In summary, we have shown that the ferromagnetic phase extends down
%% to small enough scattering lengths where, as opposed to the case
%% without optical lattice, three-body recombination does not destroy
%% the phase. Experimental verification of our results will be solid
%% confirmation of the Stoner model of itinerant ferromagnetism.

We thank S. Pilati, W. Ketterle, A. Volosniev and L. Tarruell for
discussions. Simulations were performed on the Swiss Center for
Scientific Computing (CSCS) cluster Monte Rosa in Lugano, Switzerland
and the Brutus cluster at ETH Zurich. This work was supported by ERC
Advanced Grant SIMCOFE, the Swiss National Competence Center in
Research QSIT, and the Aspen Center for Physics under grant number NSF
1066293.

\bibliographystyle{apsrev4-1}
\bibliography{PairFormation}

\end{document}